\documentclass[%
 reprint,%
 amssymb, amsmath,%
 prc,cha,%
superscriptaddress
]{revtex4}
\usepackage[colorlinks=true,linkcolor=blue]{hyperref}%
\usepackage{physics}
\usepackage{graphicx}
\usepackage{lipsum}
\usepackage{mwe}
\usepackage{caption}
\usepackage{color}

\newcommand{\sat}{\mathrm{sat}}
\newcommand{\sym}{\mathrm{sym}}
\newcommand{\expe}{\mathrm{exp}}
\newcommand{\ch}{\mathrm{ch}}
\newcommand{\isgmr}{\mathrm{ISGMR}}

\begin{document}

\title{Examining the accessibility of the compressibility of nuclear matter from nuclear experiments}

\author{J. Margueron}
\affiliation{International Research Laboratory on Nuclear Physics and Astrophysics, Michigan State University and CNRS, East Lansing, MI 48824, USA} 

\author{E. Khan}
\affiliation{Universit\'e Paris-Saclay, CNRS/IN2P3, IJCLab, 91405 Orsay, France}
\affiliation{Institut Universitaire de France (IUF)}

\begin{abstract}
The most accurate approach to determine the compressibility of nuclear matter remains the one based on microscopic Energy Density Functionals (EDFs). Recent analyses yield a value for nuclear incompressibility modulus $K_\sat=240\pm 20$~MeV, defined in nuclear matter as the second derivative of the energy per particle at saturation density. However, we demonstrate that the compressibility modulus can be reduced to values shifted by four times the suggested uncertainty, i.e., $K_\sat\approx 160$~MeV, by providing examples based on models where the second derivative ($K_\sat$) and third derivative ($Q_\sat$) of the energy per particle at saturation density can be independently varied, while the experimental binding energies, charge radii, and ISGMR data in $^{120}$Sn and $^{208}$Pb are enforced. The present work suggests a new methodology to access the compressibility of nuclear matter from nuclear experiments, still based on microscopic models, but using EDFs containing more flexibility than the ones employed up to now. Consequences of our results for nuclear matter at supra-saturation density are also discussed by exploring the quarkyonic cross-over. We predict that, for our models with low values for $K_\sat$, the quark onset density has to be low for neutron stars to exist.
\end{abstract}

\date{\today}

\maketitle

\section{Introduction}
\label{sec:intro}

The knowledge of the equation of state (EoS) for dense nuclear matter is instrumental for the understanding of astrophysical signals emitted from neutron stars (NSs) as well as gravitational waveforms originating from the fusion of two compact stars, see for instance Refs.~\cite{Takami:2014,Agathos:2015}. For symmetric matter (SM), an accurate probe of the density dependence of the energy per particle is the Isoscalar Giant Monopole Resonance (ISGMR) since it is known to be related to the incompressibility modulus of nuclear matter $K_{\sat}$ as analyzed in Ref.~\cite{Blaizot:1980} for instance. Although not directly observable, the value of $K_{\sat}$ has been used as a constraint to elaborate specific Energy Density Functional (EDF) models. Originally suggested to be around 210~MeV in the 80's~\cite{Blaizot:1980}, it was then pointing towards larger values ($260-270$~MeV) for analyses based on the relativistic mean field approach~\cite{Lalazissis:2009}, and the most recent suggested value from Ref.~\cite{Garg:2018} is now $K_\sat=240\pm20$~MeV. However, the lack of precision in the final value of $K_\sat$ may pinpoint a caveat in the method. 

Indeed, the relation between the ISGMR measurement in nuclei and the value of $K_{\sat}$ in uniform matter is not straightforward. It was investigated in several studies, see for instance the additional works in Refs.~\cite{Pearson:1991,Rudaz:1992,Vretenar:2003,Agrawal:2003,Colo:2004,Piekarewicz:2007,TLi:2007,JLi:2008,Khan:2009,Vesely:2012,WChen:2014}. Moreover, it has been underlined that the density-dependence of the nuclear incompressibility impacts its determination at saturation density~\cite{Blaizot:1995,Khan:2012,Khan:2013,Margueron:2019}. To picture this, it could be recalled that in $^{208}$Pb, the majority of nucleons (about 2/3 of the total) are localized at its surface, i.e., at sub-saturation density as shown for instance in Ref.~\cite{Khan:2012}. Therefore, nuclear experimental properties are influenced by the properties of nuclear matter below saturation density. Although this qualitative argument is to be taken with caution,  it suggests that the ISGMR strength in $^{208}$Pb may not only probe the incompressibility at saturation density $\rho_{\sat}\approx 0.155$~fm$^{-3}$, see Refs.~\cite{Margueron:2018a,Drischler:2024}, but also the density dependence $K(\rho)$ of the incompressibility from the saturation density to about 2/3 of its value, see Refs.~\cite{Khan:2012,Khan:2013} for more details. 

The present work is organized in the following way: the nuclear empirical parameters $K_\sat$ and $Q_\sat$ are introduced and discussed in Sec.~\ref{sec:nep}, then the theoretical predictions for the ISGMR are discussed in Sec.~\ref{sec:isgmr}, and the comparison between theoretical predictions and experimental measurements is shown in Sec.~\ref{sec:exp}. Finally, we discuss the consequences of our analysis for the dense matter EoS and the possible onset of a phase transition at low density. Conclusions and outlooks are given in Sec.~\ref{sec:conclusions}.

\section{The nuclear empirical parameters $K_\sat$ and $Q_\sat$}
\label{sec:nep}

\begin{figure*}[tb]
\scalebox{0.8}{\includegraphics{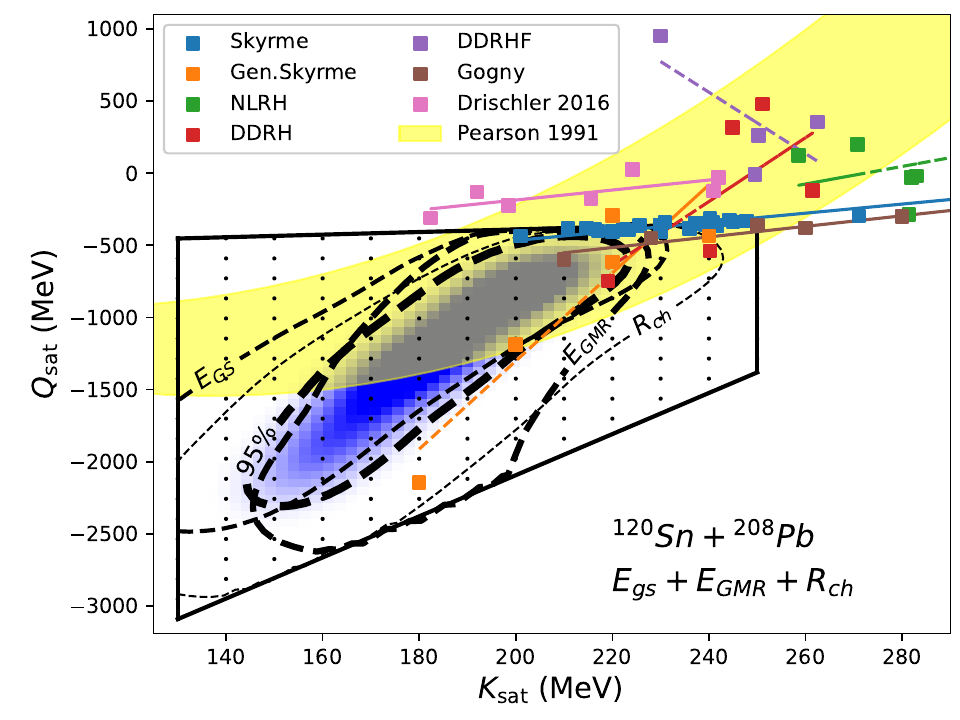}}
\caption{Domain in ($K_{\sat}$, $Q_{\sat}$) parameter space for the EDFs (squares and lines): Skyrme, Generalized Skyrme, RMF (NLRHF and DDRH), DDRHF, Gogny, MBPT 2016  and Fayans EDFs~\cite{Reinhard:2017,Miller:2019,Wang:2024} (see text for details). The parameter space explored in the present study is shown in blue surface, while the 95\% C.I. associated with the ground-state energy ($E_{gs}$), the charge radius ($R_{ch}$) and the ISGMR ($E_{GMR}$) are shown in dashed lines. The correlation suggested by Pearson~\cite{Pearson:1991} is shown in yellow.}
\label{fig:KQ:global}
\end{figure*}

At lowest order, the density dependence of the compressibility modulus is driven by the skewness parameter $Q_{\sat}$, defined as the third derivative of the energy per particle at saturation density, see Ref.~\cite{Margueron:2018a} for more details, although more dedicated quantities could be considered, such as the derivative of the incompressibility at the density $\rho_c=0.11$~fm$^{-3}$, the so-called parameter $M_c=3 \rho_c K^\prime(\rho)$ introduced in Ref.~\cite{Khan:2013}. Therefore, ISGMR measurements shall be correlated with at least two properties characterizing nuclear matter and reflecting the densities in nuclei. The simultaneous role of both $K_{\sat}$ and $Q_{\sat}$ is illustrated by their correlation, as in Fig.~5 of Ref.~\cite{Khan:2013} and reproduced in Fig.~\ref{fig:KQ:global} (colored squares) of the present paper. Results from various Skyrme, Generalized Skyrme~\cite{Farine:1997}, Gogny, Fayans~\cite{Reinhard:2017,Miller:2019,Wang:2024} EDFs, Many-Body Perturbation Theory (Drischler 2016) based on $\chi$EFT interactions calculated by Drischler~\cite{Drischler:2016} and relativistic EDFs are shown. The strong linear correlation found for each family of EDF, such as Skyrme, Fayans, and Gogny EDFs, shown with solid lines, means that setting $K_{\sat}$ defines $Q_{\sat}$. This correlation is shown in Fig.~\ref{fig:KQ:global} by the alignment of Skyrme EDFs (similarly for generalized Skyrme, Fayans, Gogny, and $\chi$EFT EDFs). This strong correlation freezes the density dependence of the incompressibility, forbidding the free exploration of ($K_{\sat}$, $Q_{\sat}$) parameter space. For the other EDFs, i.e., Density-Dependent Relativistic Hartree (DDRH) first introduced in Ref.~\cite{Typel:1999}, Non-Linear Relativistic Hartree (NLRH) suggested by~\cite{Boguta:1977}, and Density-Dependent Relativistic Hartree-Fock (DDRHF) by~\cite{Long:2006}, the correlation is weaker (dashed lines), but the explored domain of ($K_{\sat}$, $Q_{\sat}$) parameter space remains quite narrow in the figure. Note that $Q_\sat$ is identical to the parameter $S$ discussed in Refs.~\cite{Pearson:1991,Rudaz:1992,Blaizot:1995}. This parameter is sometimes called the anharmonicity parameter, since it measures the deviation from the harmonic behavior around saturation density given by $K_\sat$. 

A broader exploration of ($K_{\sat}$, $Q_{\sat}$) parameter space shall be considered to better investigate the link between available nuclear data and the density dependence of the EoS around saturation density. It is therefore important to break the tight correlation between $K_{\sat}$ and $Q_{\sat}$ observed so far, see Fig.~\ref{fig:KQ:global} for instance. For this purpose, Skyrme EDFs with extended density dependence have been considered in the present analysis. Skyrme EDFs indeed provide a simple EDF model that describes a large variety of nuclear properties at low energy, see for instance Ref.~\cite{Bender:2003}, allowing clear and efficient interpretation of nuclear data. However, the purpose of the present paper is not to suggest an extension of Skyrme EDFs but instead to investigate models breaking the correlation between $K_{\sat}$ and $Q_{\sat}$. These models provide counterexamples of the usual ones, and it will be shown in this analysis that models with compressibility modulus lower than the common expectation are possible.

\section{Theoretical predictions for the iso-scalar giant monopole resonance (ISGMR) strength distribution}
\label{sec:isgmr}

The predictions for the ISGMR, using fully microscopic approaches based on Skyrme EDFs, are performed within Constrained-HFB (CHFB) and QRPA approaches. The CHFB method, suggested in Ref.~\cite{Bohigas:1979}, is known to provide an accurate prediction of ISGMR centroid as found in Refs.~\cite{Capelli:2009,Khan:2013} by using sum rules. It allows for fast calculations, which is an advantage for systematic calculations, as in the present analysis. The ISGMR energy is calculated as 
\begin{equation}
E_{\rm ISGMR}=\sqrt{\frac{m_1}{m_{-1}}}.
\label{eq:egmr}
\end{equation} 
where $k$-th energy-weighted sum rule is defined as 
\begin{equation}
m_k=\sum_i(E_i)^k|\langle i|\hat{Q}|0\rangle |^2,
\end{equation}
with $E_i$ the RPA excitation energy and $\hat{Q}=\sum_{i=1}^A r_i^2$ the isoscalar monopole transition operator.

The $m_1$ moment is calculated from the double commutator, using the Thouless theorem~\cite{Thouless:1961}, namely $m_1=2(\hbar^2/m)A \langle r^2 \rangle$, where $A$ is the number of nucleons, $m$ nucleon mass, and $\langle r^2 \rangle$ rms radius evaluated for ground-state density provided by the considered EDF. The CHFB approach has been considered in Refs.~\cite{Bohigas:1979,Blaizot:1995,Capelli:2009,Khan:2009} to obtain the $m_{-1}$ moment. It consists in adding to EDF hamiltonian $\hat{H}_\text{EDF}$ a constraint associated with isoscalar monopole operator, namely $\hat{H}_\text{constr}=\hat{H}_\text{EDF}+\lambda\hat{Q}$. The $m_{-1}$ moment is then obtained from the derivative of the expectation value of the monopole operator on CHFB solution $\vert\lambda\rangle$,
\begin{equation}
m_{-1}=-\frac{1}{2}\left[\frac{\partial}{\partial\lambda}\langle\lambda|
\hat{Q}|\lambda\rangle\right]_{\lambda=0}.
\end{equation}

\begin{table*}[t]
\tabcolsep=0.3cm
\def\arraystretch{1.4}
\begin{tabular}{cccccccccccc}
\hline\hline
$K_\sat$ & $Q_\sat$ & CHFB & QRPA & $t_{0}$ & $x_{0}$ & $t_{3}$ & $x_{3}$ & $\gamma$ & $t_{3}^\prime$ & $x_{3}^\prime$ & $\gamma^\prime$ \\
MeV & MeV & MeV & MeV \\
\hline
\multicolumn{12}{c}{$^{120}$Sn with $E_{\isgmr}^{\expe}=15.5\pm 0.2$~MeV from \cite{liga:2010}}\\
\hline
180 & -1491 & 15.63 & 15.40 & -1287 & 0.258 & 14634 & 1.04 & 0.80 & -34183 & 0.666 & 2.40 \\ 
180 & -1412 & 15.63 & 15.39 & -1264 & 0.235 & 21977 & 0.986 & 0.95 & -33679 & 0.747 & 1.90 \\
210 & -470 & 15.66 & 15.58 & -2162 & 0.585 & 13739 & 1.028 & 0.25 & -2998 & 0.196 & 1.00 \\
210 & -522 & 15.75 & 15.65 & -1840 & 0.511 & 13281 & 1.040 & 0.35 & -4926 & 0.537 & 1.05 \\
\hline
\multicolumn{12}{c}{$^{208}$Pb with $E_{\isgmr}^{\expe}=13.5\pm 0.2$~MeV from \cite{uchida:03}}\\
\hline
180 & -1531 & 13.23 & 13.34 & -1320 & 0.284 & 12270 & 1.071 & 0.70 & -41580 & 0.596 & 2.80 \\
180 & -1895 & 13.44 & 13.58 & -1206 & 0.210 & 16057 & 1.074 & 0.95 & -63869 & 0.747 & 2.85 \\
210 & -418 & 13.18 & 13.25 & -2587 & 0.589 & 17180 & 0.860 & 0.20 & -3449 & -0.131 & 0.60 \\
210 & -540 & 13.36 & 13.43 & -1708 & 0.462 & 16439 & 1.000 & 0.45 & -8964 & 0.713 & 0.90 \\
\hline\hline
\end{tabular}
\caption{Comparison of the centroid energies defined as $E_{\isgmr}=\sqrt{m_1/m_{-1}}$ obtained for CHFB and QRPA approaches for several extended Skyrme EDFs for which the parameters $t_{0}$, $x_{0}$, $t_{3}$, $x_{3}$, $t_{3}^\prime$, $x_{3}^\prime$, $\gamma$ and $\gamma^\prime$ are explicitly given. The other Skyrme parameters are the same as for SLy5, i.e., $t_1=483.13$, $x_1=-0.328$, $t_2=-549.4$, and $x_2=-1.0$.}
\label{tab:compare}
\end{table*}

In addition to the CHFB approach, the QRPA has the advantage of providing the strength function, see for instance the spherical Skyrme-QRPA model presented in Ref.~\cite{colo:2021}, where the whole residual interaction is considered and the QRPA equations are expressed in matrix form involving the A and B matrices~\cite{RingSchuck:1980}:
\begin{eqnarray}
\left(
\begin{array}{c c}
A_{\alpha\beta,\gamma\delta}         & B_{\alpha\beta,\gamma\delta}         \\
-B^{\ast}_{\alpha\beta,\gamma\delta} & -A^{\ast}_{\alpha\beta,\gamma\delta} \\
\end{array}
\right)
\left(
\begin{array}{c}
X^{\nu}_{\gamma\delta}  \\ Y^{\nu}_{\gamma\delta} \\
\end{array}
\right)
&=&
E_{\nu}
\left(
\begin{array}{c}
X^{\nu}_{\alpha\beta}  \\ Y^{\nu}_{\alpha\beta} \\
\end{array}
\right).
\end{eqnarray}

We first assess the compatibility between CHFB and QRPA for a few extended Skyrme EDFs. A comparison between the CHFB and the QRPA for $E_{\isgmr}$ is given in Tab.~\ref{tab:compare} for $^{120}$Sn and $^{208}$Pb and for a set of four extended Skyrme EDFs among those considered in the present work, as described below. It illustrates that these two approaches provide values that are comparable within a theoretical uncertainty of $\approx 0.1$-$0.2$~MeV. We obtained similar results with other extended Skyrme EDFs given in App.~\ref{ap:GSkyrme}.

The description of the ISGMR in both Sn and Pb nuclei is known to be one of the most stringent tests for EDFs. For instance, Refs.~\cite{Litvinova:2023,ZLi:2023} have recently shown that considering configurations mixing beyond 1particle-1hole improves the theoretical description of this data. Therefore, this data shall be considered, among others, to design new EDFs with extended density dependence.

For our demonstration, a term of the following form, $\frac{1}{6}t^\prime_3(1+x^\prime_3P_\sigma)\rho^{\gamma^\prime}\delta(\vec{r}_1-\vec{r}_2)$, where $t^\prime_3$, $x^\prime_3$ and $\gamma^\prime$ are new parameters and P$_\sigma$ is the spin exchange operator, has been added to standard Skyrme EDFs. This new term can be justified as representing better the in-medium many-body forces than standard Skyrme EDFs. It generates contributions to both the Hartree-Fock field and the residual interaction, which are analogous to the ones generated by the usual density-dependent interaction, see Ref.~\cite{Dobaczweski:1984}. 

In the case of the pairing channel (which matters for $^{120}$Sn), we consider a mixed contact pairing interaction of the following form: $300\hbox{ MeV~fm$^{-3}$}[1-0.5(\rho/\rho_\sat)^{0.2}]\delta(r)$. This mixed pairing form impacts mostly the binding energy in $^{120}$Sn. The parameter 0.5 influences the mass-number dependence of the pairing gap: an intermediate value between the two extremes (0 and 1) is adopted in Refs.~\cite{Dobaczewski:2001} and \cite{Dobaczewski:2002}. The power of density, 0.2, may affect the appearance of neutron skins and halos as shown by~\cite{Dobaczewski:2001}. The values of the pairing interaction parameters are fixed to get a pairing gap of about 1.25~MeV in $^{120}$Sn with SLy4 Skyrme interaction. Note that the pairing interaction could also be completed with isovector terms, see Refs.~\cite{Margueron:2008} and \cite{Yamagami:2012}. For the calculation of charge radii, we consider spin-orbit and Darwin-Foldy contributions, which provide magnetic dipole moment and relativistic corrections to nuclear rms charge radius as shown by~\cite{Bender:2003} and \cite{Carlson:2023}.

In this framework, nuclear empirical parameters (NEP) $K_\sat$ and $Q_\sat$ in symmetric matter (SM) are expressed in the following way
\begin{eqnarray}
K_\sat &=& \left(
-\frac{3}{5} \frac{\hbar^2}{2m} 
+6 C_0^\tau\rho_\sat \right) \left(\frac{3\pi^2}{2}\right)^{2/3}\rho_\sat^{2/3} \nonumber \\
&&+9\left(2C_0^{\prime} + C_0^{\prime\prime}\rho\right)\rho_\sat^2 \\
Q_\sat &=& \left(
\frac{12}{5} \frac{\hbar^2}{2m} 
-6 C_0^\tau\rho_\sat \right) \left(\frac{3\pi^2}{2}\right)^{2/3}\rho_\sat^{2/3} \nonumber \\
&&+27\left(3C_0^{\prime\prime} + C_0^{\prime\prime\prime}\rho\right)\rho_\sat^3 \, ,
\end{eqnarray}
with $C_0^\tau = \left[ 3t_1+(5+4x_2)t_2\right]/16$ and
\begin{eqnarray}
C_0^{\prime} &=& \frac{1}{16}\Big[ \gamma t_3\rho_\sat^{\gamma-1} + \gamma^\prime t^\prime_3\rho_\sat^{\gamma^\prime-1} \Big]\, ,\\
C_0^{\prime\prime} &=& \frac{1}{16}\Big[ \gamma (\gamma-1) t_3\rho_\sat^{\gamma-2} + \gamma^\prime (\gamma^\prime-1) t^\prime_3\rho_\sat^{\gamma^\prime-2} \Big] \, ,\\
C_0^{\prime\prime\prime} &=& \frac{1}{16}\Big[ \gamma (\gamma-1)(\gamma-2) t_3\rho_\sat^{\gamma-3} \nonumber \\
&&\hspace{0.5cm}+ \gamma^\prime (\gamma^\prime-1)(\gamma^\prime-2) t^\prime_3\rho_\sat^{\gamma^\prime-3} \Big]\, ,
\end{eqnarray}
where combinations of these parameters are fixed by the saturation density $\rho_\sat$, the binding energy at saturation $E_\sat$, and the effective mass $m^*_\sat/m$. Two additional parameters in SM, namely $t_3^\prime$ and $\gamma^\prime$ -- since $x_3^\prime$ impacts only asymmetric matter -- allow us to widely explore the NEPs $K_\sat$ and $Q_\sat$, and to break the linear correlation among them shown in Fig.~\ref{fig:KQ:global}. Actually, large uncertainties for $Q_{\sat}$, about $\pm$ 1 GeV, has been suggested in Refs.~\cite{Khan:2013,Margueron:2018a,Grams:2022} and we explore a large range of values for $K_{\sat}$, well outside the usual 220-260~MeV domain. However, to limit the choice of parameters, we impose the same isovector NEPs of the SLy5 Skyrme EDF, namely the symmetry energy $E_\sym$, and its density dependence $L_\sym$. In the future, a wider domain of NEPs can be explored by releasing this limitation.

\section{Comparison between theoretical predictions and experimental measurements}
\label{sec:exp}

We consider the following three sets of experimental constraints: ISGMR centroid energy $E_{\isgmr}^{\expe}(^{120}$Sn$)= 15.5\pm$0.2~MeV from Ref.~~\cite{liga:2010} and $E_{\isgmr}^{\expe}(^{208}$Pb$)=13.5\pm$0.2~MeV from Ref.~\cite{uchida:03}; binding energy $B^{\expe}(^{120}$Sn$)=-1020\pm5$~MeV and $B^{\expe}(^{208}$Pb$)=-1636\pm2$~MeV from Ref.~\cite{wang:21}; and charge radii are $R_{\ch}^{\expe}(^{120}$Sn$)=4.65\pm0.02$~fm and $R_{\ch}^{\expe}(^{208}$Pb$)=5.50\pm0.01$~fm from Ref.~\cite{taoli:21}. Systematic uncertainties are considered to fix the error bars. A likelihood probability $p=\exp(-\chi^2/2)$ is associated with each model, where $\chi^2$ reflects the above-mentioned experimental constraints, see for instance recent Ref.~\cite{Giuliani:2023} for details on the Bayesian approach. For each of these three sets of constraints, the 95\% confidence interval is shown in Fig.~\ref{fig:KQ:global} with different dashed lines. The thick dashed line in Fig.~\ref{fig:KQ:global} is obtained by combining all three constraints: it delimits 95\% of the probability distribution (shown in blue where deeper blue means larger probability). In the future, it would be interesting to analyze the impact of the surface energy (defined in semi-infinite matter) on the barrier heights in deformed nuclei and fission processes.

The key feature in our demonstration is that the new models, describing ISGMR, charge radii, and binding energies in both $^{120}$Sn and $^{208}$Pb, explore wide values for $Q_{\sat}$ and $K_{\sat}$, as it can be seen in Fig.~\ref{fig:KQ:global}. The ($K_{\sat}$, $Q_{\sat}$) correlation has been broadened, as aimed, namely the blue area compared to the squares representing EDF predictions. Note also the large agreement between our blue region and the yellow correlation suggested in Ref.~\cite{Pearson:1991} and reported in Ref.~\cite{Rudaz:1992}. The yellow correlation is based on an empirical analysis of both experimental and theoretical results, and it is extrapolated to low values for $K_\sat$ that the original authors have not explored. The blue region shows that it is possible to obtain counterexamples exploring regions for the parameters $K_{\sat}$ and $Q_{\sat}$ that are unexplored by other approaches.

Extensions of standard Skyrme EDF have been suggested in the literature, see for instance Refs.~\cite{Farine:1997,Lesinski:2006,Chamel:2009,Erler:2010,Gil:2019} and references therein. It is interesting to note that a model with a low value for $K_\sat\approx 180$~MeV has been obtained in Ref.~\cite{Farine:1997} but for a very specific value of $Q_{sat}$ that makes this EDF out of the 95 \% probability distribution (blue) domain, see Fig.~\ref{fig:KQ:global}. Models with values for $K_\sat$ as low as 180~MeV have also been obtained in Ref.~\cite{Erler:2010} by fitting only nuclear ground state properties (without the ISGMR). However, the authors have not calculated the value of $Q_\sat$ associated with their values for $K_\sat$, so we have not added these extensions to our analysis. Table~\ref{tab:compare} presents a set of extended Skyrme parameterizations obtained in the present analysis, and the complete list of models is given in App.~\ref{ap:GSkyrme}. 

\begin{figure*}[tb]
\scalebox{0.6}
{\includegraphics{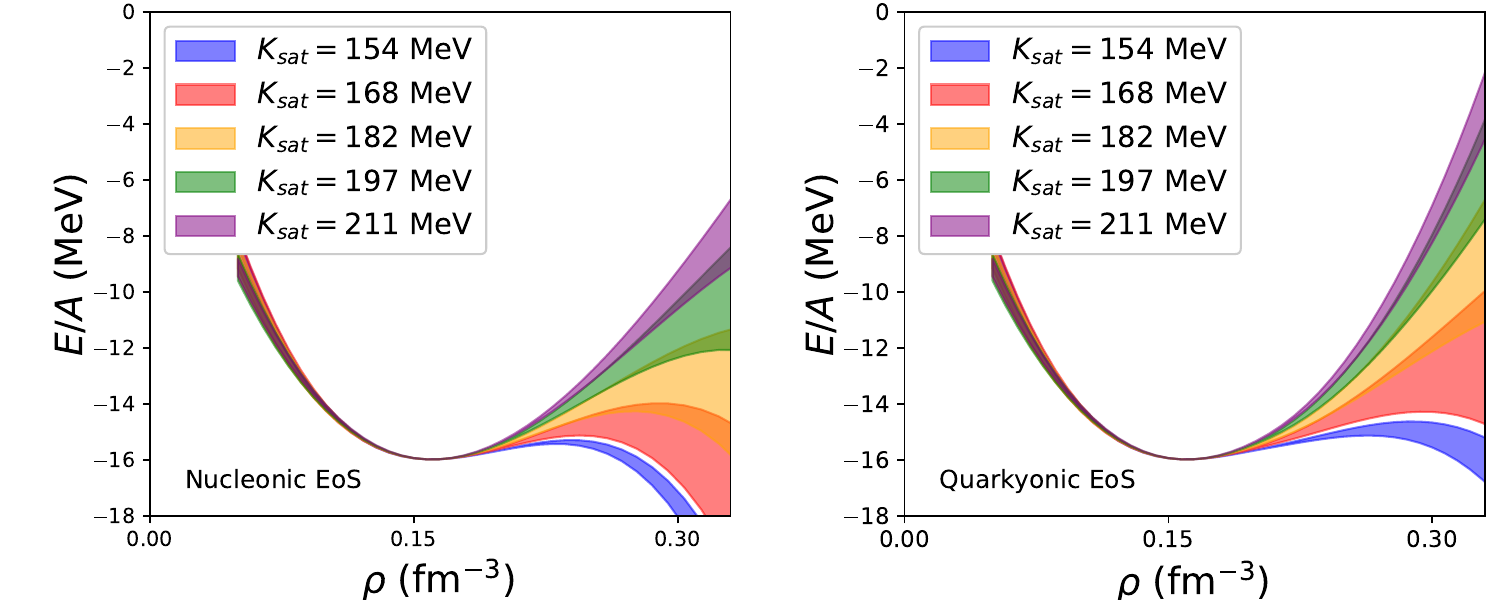}}
\caption{Binding energy $E/A$ for the nuclear equation of state based on the EDFs shown in Fig.~\ref{fig:KQ:global}. left: nucleonic models, right: quarkyonic models.}
 \label{fig:eos}
\end{figure*}

\section{Prediction for the equation of state above saturation density}
\label{sec:eos}

Let us investigate the behavior of the EoS with low values for $K_\sat$ existing in the blue region of Fig.~\ref{fig:KQ:global}. These EoS are shown in Fig.~\ref{fig:eos} for SM as a function of the particle density. The range of $Q_\sat$ values associated with a value for $K_\sat$ in the validated domain (blue area on Fig. \ref{fig:KQ:global}) is treated as uncertainties, giving rise to bands on the EoS. We also limit our density exploration to about 2$\rho_\sat$, since the properties of nuclei provide constraints only around $\rho_\sat$ and not very far above. It should be noted that nucleonic models with low $K_\sat$ values ($\lesssim$ 180 MeV), explore also low $Q_\sat$ values, as shown in Fig.~\ref{fig:KQ:global} (left). These models also predict that supra-saturation matter is more bound than matter at saturation density (nuclear matter collapse). However, a phase transition or a crossover, like the one provided by the quarkyonic model~\cite{McLerran:2019} and extended in asymmetric matter in Ref.~\cite{Margueron:2021}, can modify this behavior above $\rho_\sat$ without impacting finite nuclei. Such an example of the quarkyonic model with a smooth transition to quark matter starting at 0.2~fm$^{-3}$ is shown in the right panel of Fig.~\ref{fig:eos}: a transition to quark matter starting above saturation density -- therefore not modifying finite nuclei properties -- could stabilize purely nucleonic models with low values of $K_\sat$, except for $K_\sat \lesssim$ 160 MeV.

\section{Conclusions and outlooks}
\label{sec:conclusions}

In conclusion, it is shown that the uncertainty in $K_{\sat}$ obtained in recent data analyses, see for instance Ref.~\cite{Colo:2004} and references therein, is underestimated. Examples are shown, see Fig.~\ref{fig:KQ:global}, which predict values for $K_\sat$ up to four sigma away from the standard one and still describe the ISGMR, binding energies, and charge radii data. While this result is based on well-known techniques, e.g., CHFB and RPA as described above, the existence of these examples questions the common methodologies employing over-constrained models -- with correlations among parameters -- to analyze data in finite nuclei, like the ISGMR energy in $^{120}$Sn and $^{208}$Pb. These correlations are most probably artificial since these EDFs have been constructed by reducing as much as possible the number of parameters, as it is common in nuclear modeling. These induced correlations among parameters introduce a bias in the analysis of experimental data by limiting the free exploration of the parameter space. We suggest, instead, to analyze experimental data for ISGMR energies by employing flexible models where the parameters $K_\sat$ and $Q_\sat$ can be widely and independently varied. It is the condition for the uncertainties in $K_\sat$ and $Q_\sat$ to reflect those of the data, and not the internal correlations within the model.

In the case of the strong softening of nuclear matter above the saturation density suggested by some of our models and leading to nuclear matter collapse, a mechanism producing repulsion is necessary to stabilize neutron stars. We have explored one of them in our analysis: a low-density crossover between nuclear and quark matter provided by the quarkyonic model~\cite{McLerran:2019}. The present results show that precise measurements and accurate analyses of finite nuclear properties, such as, for instance, the ISGMR, can impact the onset of a new phase in the core of neutron stars.

The data that support the findings of this article are given in Tables~\ref{tab:compare} and \ref{tab:fulllist}.
Fig.~\ref{fig:KQ:global} can be generated using the Python toolkit nucleardatapy available on Pipy and GitHub, for more details see Ref.~\cite{Margueron:2026}.

\section*{Acknowledgements}
We thank J.-P. Blaizot, W. Nazarewicz, and M. Urban for valuable discussions, and P.-G. Reinhard for providing us the values for $K_\sat$ and $Q_\sat$ predicted by Fayans EDFs and shown in Fig.~1. J.M. and E.K. are supported by the CNRS-IN2P3 MAC masterproject, and this work benefited from the support of the project RELANSE ANR-23-CE31-0027-01 of the French National Research Agency (ANR). J.M. benefits from the LABEX Lyon Institute of Origins (ANR-10-LABX-0066). 

\appendix

\section{Selected parameter sets}
\label{ap:GSkyrme}

The complete list of extended Skyrme EDF parameter sets compatible with the experimental binding energies, charge radii, and ISGMR data in $^{120}$Sn and $^{208}$Pb is given in Table~\ref{tab:fulllist}.

\begin{table*}[t]
\tabcolsep=0.3cm
\def\arraystretch{1.4}
\begin{tabular}{ccccccccc}
\hline\hline
name & $t_0$ & $x_0$ & $t_3$ & $x_3$ & $t_3^\prime$ & $x_3^\prime$ & $\gamma$ & $\gamma^\prime$ \\
\hline
\hline
SGMR160-3-75 & -1343.43 & 0.2624 & 15081.47 & 0.9475 & -32050.13 & 0.4905 & 0.75 & 2.25 \\ 
SGMR160-3-80 & -1306.21 & 0.2463 & 15382.97 & 0.9652 & -38848.65 & 0.5260 & 0.80 & 2.40 \\ 
SGMR160-3-85 & -1273.67 & 0.2309 & 15764.64 & 0.9810 & -47359.56 & 0.5584 & 0.85 & 2.55 \\
SGMR170-2-75 & -1413.99 & 0.2853 & 20480.78 & 0.8646 & -22836.32 & 0.5570 & 0.75 & 1.50 \\
SGMR170-2-85 & -1335.11 & 0.2561 & 21367.45 & 0.9062 & -28266.65 & 0.6229 & 0.85 & 1.70 \\
SGMR170-2-90 & -1302.71 & 0.2422 & 21970.49 & 0.9239 & -31684.81 & 0.6514 & 0.90 & 1.80 \\
SGMR170-3-80 & -1296.59 & 0.2518 & 15008.60 & 1.0017 & -36515.99 & 0.5915 & 0.80 & 2.40 \\ 
SGMR170-3-85 & -1265.15 & 0.2358 & 15401.25 & 1.0160 & -44640.54 & 0.6229 & 0.85 & 2.55 \\
SGMR170-3-90 & -1237.40 & 0.2206 & 15867.61 & 1.0287 & -54833.49 & 0.6514 & 0.90 & 2.70 \\
SGMR180-2-95 & -1263.71 & 0.2347 & 21976.53 & 0.9857 & -33678.46 & 0.7469 & 0.95 & 1.90 \\
SGMR180-3-65 & -1407.54 & 0.3129 & 13884.45 & 0.9923 & -19172.66 & 0.5551 & 0.65 & 1.95 \\
SGMR180-4-50 & -1543.63 & 0.3746 & 12169.66 & 0.9986 & -14463.65 & 0.3938 & 0.50 & 2.00 \\
SGMR180-4-60 & -1412.33 & 0.3269 & 12028.10 & 1.0402 & -24051.70 & 0.5084 & 0.60 & 2.40 \\ 
SGMR180-4-70 & -1320.07 & 0.2840 & 12270.04 & 1.0705 & -41580.33 & 0.5964 & 0.70 & 2.80 \\
SGMR190-2-75 & -1381.15 & 0.3039 & 18925.97 & 0.9768 & -19769.11 & 0.7210 & 0.75 & 1.50 \\
SGMR190-2-85 & -1309.55 & 0.2709 & 19913.86 & 1.0088 & -24823.25 & 0.7786 & 0.85 & 1.70 \\
SGMR190-3-45 & -1672.91 & 0.4129 & 13752.03 & 0.9469 & -8966.23 & 0.4213 & 0.45 & 1.35 \\
SGMR190-3-55 & -1506.49 & 0.3656 & 13351.30 & 1.0015 & -12303.57 & 0.5523 & 0.55 & 1.65 \\
SGMR190-4-65 & -1351.56 & 0.3114 & 11821.12 & 1.0943 & -28949.41 & 0.6480 & 0.65 & 2.60 \\
SGMR190-4-70 & -1310.65 & 0.2897 & 11998.58 & 1.1060 & -38412.78 & 0.6868 & 0.70 & 2.80 \\
SGMR200-2-45 & -1753.62 & 0.4372 & 17685.93 & 0.8949 & -10384.39 & 0.5474 & 0.45 & 0.90 \\
SGMR200-2-60 & -1509.82 & 0.3729 & 17320.35 & 0.9852 & -13284.75 & 0.7185 & 0.60 & 1.20 \\ 
SGMR200-2-80 & -1328.60 & 0.2956 & 18622.84 & 1.0538 & -20481.97 & 0.8503 & 0.80 & 1.60 \\
SGMR200-2-90 & -1268.50 & 0.2622 & 19839.29 & 1.0768 & -26152.43 & 0.8949 & 0.90 & 1.80 \\ 
SGMR200-4-50 & -1506.68 & 0.3961 & 11431.68 & 1.0953 & -11591.73 & 0.6152 & 0.50 & 2.00 \\
SGMR210-2-50 & -1618.86 & 0.4367 & 16281.96 & 1.0281 & -9773.91 & 0.7729 & 0.50 & 1.00 \\ 
SGMR210-2-65 & -1432.05 & 0.3653 & 16626.77 & 1.0843 & -13258.10 & 0.8932 & 0.65 & 1.30 \\
SGMR210-2-75 & -1348.31 & 0.3235 & 17371.16 & 1.1089 & -16701.90 & 0.9453 & 0.75 & 1.50 \\
\hline\hline
\end{tabular}
\caption{Complete list of parameter sets. The other Skyrme parameters are the same as for SLy5, i.e., $t_1=483.13$, $x_1=-0.328$, $t_2=-549.4$, and $x_2=-1.0$.}
\label{tab:fulllist}
\end{table*}

\bibliography{biblio}

\end{document}